\documentclass[twocolumn,showpacs,showkeys,preprintnumbers,amsmath,amssymb,superscriptaddress]{revtex4}

\usepackage{graphicx}
\usepackage{dcolumn}
\usepackage{bm}

\begin{document}

\title{Lifetime measurement of the 6792~keV state in $^{15}$O important for the astrophysical S~factor extrapolation in $^{14}$N(p,$\gamma$)$^{15}$O} 

\author{D.~Sch\"urmann}
	\altaffiliation[present address: ]{Deparment of Physics, University of Notre Dame, Notre Dame, Indiana 46556, USA}
	\affiliation{Institut f\"ur Experimentalphysik III,
	Ruhr-Universit\"at Bochum, Germany}
\author{R.~Kunz}
 	\affiliation{Institut f\"ur Experimentalphysik III,
	Ruhr-Universit\"at Bochum, Germany}
\author{I.~Lingner}
	\altaffiliation[present address: ]{Astronomisches Institut, Ruhr-Universit\"at Bochum, Germany}
 	\affiliation{Institut f\"ur Experimentalphysik III,
	Ruhr-Universit\"at Bochum, Germany}
\author{C.~Rolfs}
 	\affiliation{Institut f\"ur Experimentalphysik III,
	Ruhr-Universit\"at Bochum, Germany}	
\author{F.~Sch\"umann}
	\altaffiliation[present address: ]{Lessingschule, Bochum, Germany}
	\affiliation{Institut f\"ur Experimentalphysik III,
	Ruhr-Universit\"at Bochum, Germany}
\author{F.~Strieder}
	\email[corresponding author: ]{strieder@ep3.rub.de}
	\affiliation{Institut f\"ur Experimentalphysik III,
	Ruhr-Universit\"at Bochum, Germany}
\author{H.-P.~Trautvetter}
	\affiliation{Institut f\"ur Experimentalphysik III,
	Ruhr-Universit\"at Bochum, Germany}

\date{\today}

\begin{abstract}

We report on a new lifetime measurement of the $E_x = 6792$~keV state in $^{15}$O via the Doppler-shift attenuation method at the ${\rm E}  = 259$~keV resonance in the reaction $^{14}$N(p,$\gamma$)$^{15}$O. This subthreshold state is of particular importance for the determination of the ground state astrophysical S~factor of $^{14}$N(p,$\gamma$)$^{15}$O at stellar energies. The measurement technique has been significantly improved compared with previous work. The conclusion of a finite lifetime drawn there cannot be confirmed with the present data. In addition, the lifetime of the two states at $E_x = 5181$ and 6172~keV have been measured with the same technique in order to verify the experimental method. We observe an attenuation factor ${\rm F}(\tau)>0.98$ for the $E_x =6172$ and 6792~keV states, respectively, corresponding to $\tau<0.77$~fs. The attenuation factor for the $E_x =5181$~keV state results in ${\rm F}(\tau) = 0.78\pm0.02$ corresponding to $\tau=8.4\pm1.0$~fs in excellent agreement with literature.

\end{abstract}

\email{strieder@ep3.rub.de}

\maketitle

\section{Introduction}

\begin{figure}
  \includegraphics[angle=0,width=7cm]{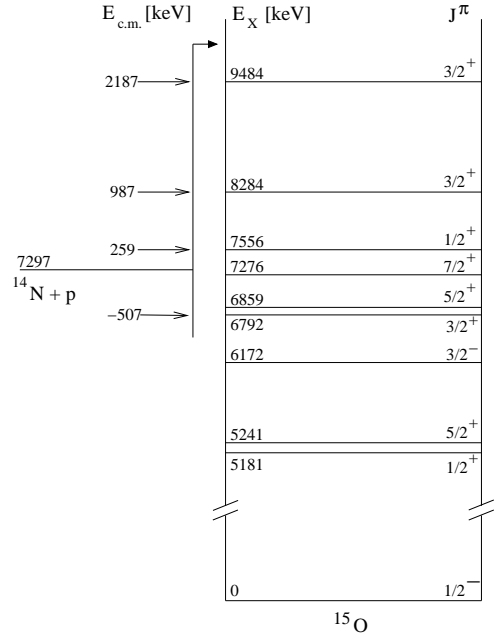}
  \caption{\label{leveldia}Level structure of $^{15}$O \cite{imb05}.}
\end{figure}

\begin{figure}
  \includegraphics[angle=0,width=\columnwidth]{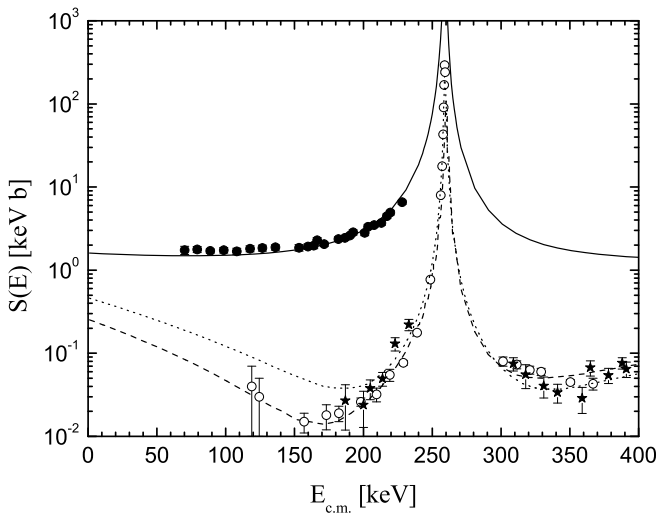}
  \caption{\label{fig1}Astrophysical S~factor of $^{14}$N(p,$\gamma)^{15}$O. Shown are the measured total S~factor (full circles, \cite{lem06}), the ground state S~factor of Formicola et al. (open circles, \cite{for04}) and Runkle et al. (stars, \cite{run05}), and the R-matrix calculation of \cite{for04} for the total S~factor (solid line) and the ground state transition (dashed line) as well as the ground state R-matrix calculation of \cite{ber01} (dotted line).}
\end{figure}

The capture reaction $^{14}$N(p,$\gamma$)$^{15}$O (${\rm Q} = 7297$~keV, fig. \ref{leveldia}) is the slowest process in the hydrogen burning CNO cycle \cite{rol88} and thus of high astrophysical interest. This reaction plays a key role for the energy production of more massive main sequence stars and the detailed understanding of the neutrino spectrum of our sun \cite{ahl90, bor07} as well as the age determination of globular cluster stars \cite{imb04}.  The reaction was recently studied in three experiments at energies ranging from $E = 70$ to 480~keV \cite{for04, imb05, run05, lem06} (all energies are given in the center-of-mass frame if not indicated different) and previously over a wide range of energies, i.e. $E = 240$ to 3300~keV (\cite{sch87} and references therein). The major motivation for reinvestigating this reaction was a reanalysis \cite{ang01} of the data of \cite{sch87} using R-matrix resulting in a significant change in the extrapolation of the ground state astrophysical S~factor from ${\rm S_{gs}}(0) = 1.65$~keV b \cite{sch87} to a nearly negligible contribution.  A reduction of the ground state contribution could indeed be verified by the recent experiments \cite{for04, run05}. However, 
the R-matrix analysis revealed that below the $E_R=259$~keV $J^\pi=1/2^+$ resonance the data followed primarily the low energy wing of the resonance (fig. \ref{fig1}) and could not probe significantly the behavior of the interference structure of the four $J^\pi=3/2^+$ resonances at $E_R=-507$, 987, 2187~keV and the background pole - here fixed at 6~MeV - around the energy of this $J^\pi=1/2^+$ resonance which in turn adds incoherently to the ground state S~factor (for details see \cite{for04} and \cite{tra08}). A precise knowledge of this interference structure is needed for reliable extrapolation due to a minimum of the S~factor curve near $E=160$~keV.

In a second experiment at the LUNA accelerator at the Laboratori Nazionali del Gran Sasso, Italy, with a 4$\pi$ BGO detector and a windowless gas target system \cite{lem06} the total S~factor of $^{14}$N(p,$\gamma)^{15}$O has been measured down to $E=70$~keV with $S(70)=1.74\pm0.14$~keV~b (statistical error only). However, the ground state contribution at that energy is expected to be  ${\rm S}_{gs}(70)=0.07$~keV~b and falls rapidly above 70~keV up to the resonance \cite{for04}. It is obvious that the experiment \cite{lem06} cannot probe sensitively the contribution of the ground state transition within a factor of two of the results of \cite{for04}. Fig. \ref{fig1} shows a comparison of the available data together with R-matrix fits \cite{for04, run05}. Note: the data of \cite{lem06} are not taken into account in the R-matrix analysis. 

The energy dependence of S$_{gs}$ at low energies is primarily controlled by the width of the subthreshold state at $E_x=6792$~keV. The width could be deduced by a measurement of the lifetime of this state. In this case such an experiment is feasible by means of $\gamma$ spectroscopy if the lifetime is long enough to allow for a observable Doppler-shift attenuation in the slowing down process of the recoiling $^{15}$O nucleus. This method is called Doppler-Shift Attenuation Method (DSAM) and was reported in \cite{ber01} with an attenuation factor $F(\tau)=0.93\pm0.03$ corresponding to a lifetime of $\tau =1.6\pm0.8$~fs. In this experiment the authors used three Ge detectors placed at 0$^\circ$, 90$^\circ$, and 144$^\circ$ with respect to the beam axis. The advantage of this approach is that all information in principal could be obtained in a single run, but results in the difficulty that each spectrum had to be calibrated absolutely and as a function of time to a high precision (better than $\pm0.5$~keV) in order to extract an $F(\tau)$ value near unity.

In contrast to the previous experiment \cite{ber01} the present approach uses one detector placed on a turntable with an angle span from -40$^\circ$ to +116$^\circ$. Gamma spectra were obtained at 11 angles of which 2 angle settings were symmetric around the beam axis leaving 9 different angles. In this way a left/right asymmetry in the detector placement could be detected as well as a possible asymmetry in the beam alignment. Therefore systematic uncertainties could be greatly reduced. The gain stability of the electronics in the course of the experiment was checked by the $E_\gamma=6129$~keV $\gamma$-ray emission from the reaction $^{19}$F(p,$\alpha\gamma)^{16}$O. Due to the fact that all measurements were done with the same detector a precise absolute determination of the $\gamma$-ray energy was not necessary. The determination of the slope of the energy calibration was obtained via the $^{14}$N(p,$\gamma)^{15}$O reaction itself using the $\gamma$-ray energy information of the various transitions in $^{15}$O given in \cite{imb05}.

\begin{figure}[b]
  \hspace{-0.4cm}
  \includegraphics[angle=90,width=8cm]{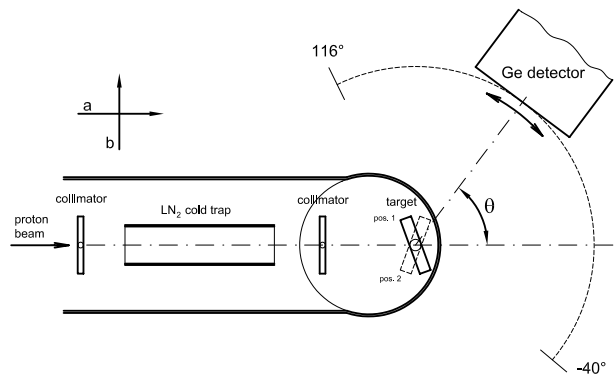}
  \vspace{-0.3cm}
  \caption{\label{figsetup}Schematic diagram of the experimental set-up. In the upper left corner the directions of the geometrical offset parameter a and b of the turntable axis are indicated (for details see text). }
\end{figure}

\section{Experimental set-up and experimental procedure}
The 500~kV accelerator at the Dynamitron tandem laboratory at the Ruhr-Universit\"at Bochum provided the proton beam with typically 100 to 150~$\mu$A on target at $E_p=318$~keV. The energy calibration and the energy spread of the proton beam are known to better than 1~keV, and equal to 70~eV, respectively \cite{klu05}. Fig. \ref{figsetup} shows a schematic diagram of the experimental set-up. The beam passed through two Ta collimators with a diameter of 10 and 4~mm at a distance to the target of 80 and 6.5~cm, respectively, and was stopped in the N target. A LN$_2$ cooled shroud placed at 20~cm in front of the target avoided C-build up and ensured a typical pressure in the target chamber of about $5\times10^{-7}$~mbar. After focusing the beam, the 4~mm collimator was removed to minimize beam induced background originating from the collimator. The target chamber was formed by a cylindrical T-shape steel tube (inner diameter = 9.7~cm, wall-thickness = 2~mm) in order to ensure the same influence of the $\gamma$-ray absorption in the chamber walls for all angles. The target was mounted on a directly water cooled, rotatable target holder. On the target holder three targets could be placed at the same time and brought into the beam without breaking the vacuum. The alignment of the target chamber and the target itself was checked to be better than $\pm1$~mm. 

A turntable was placed at the target chamber with its axis at the target position. The 0$^\circ$ position could be aligned to the beam axis within $\pm0.5^\circ$. The total uncertainty of the turntable axis position in respect to the beam spot position could be estimated by $\pm3$~mm, where the uncertainty coming from the detector positioning on the turntable arm as well as the beam spot position on the target are included. The angles reached by this set-up ranged from -40$^\circ$ to 116$^\circ$ with respect to the beam axis (fig. \ref{figsetup}) and reproducibility within less than 0.2$^\circ$ was varified. A HPGe gamma detector was placed on the arm of the turntable (108\% relative efficiency, resolution = 2.1~keV at $E_\gamma = 1.3$~MeV, front face distance to the target = 13.1~cm).

\begin{figure}
  \includegraphics[angle=0,width=\columnwidth]{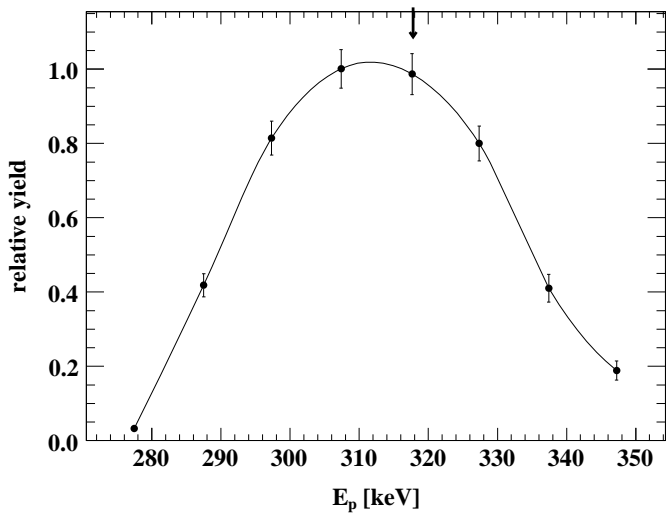}
  \caption{\label{figscan}Thick-target $\gamma$-ray yield curve of the $E_R=259$~keV resonance (proton energy $E_p=277$~keV) of $^{14}$N(p,$\gamma)^{15}$O obtained with a $^{14}$N implanted Ta sheet (220~keV $^{14}$N ion beam). The proton energy $E_p = 318$~keV for all the runs is indicated by the arrow.}
\end{figure}

The N target has been produced by implanting $^{14}$N into a Ta backing with a 0.5~mm thickness (backing diameter = 50~mm). A molecular N beam at an energy of $E_{^{14}N_2}=220$~keV from the 500~kV accelerator was rastered with a pair of magnetic x-y scanners over an aperature (diameter = 24~mm) in close geometry to the target leading to an implanted area of 25~mm diameter. The $^{14}$N distribution in the Ta was investigated using the $E_R=259$~keV resonance of $^{14}$N(p,$\gamma)^{15}$O. A target profile obtained after an $^{14}$N irradiation dose of $3.9\times10^{18}$ atoms/cm$^2$ is shown in fig. \ref{figscan}. As seen in the figure the $^{14}$N atoms are distributed from the surface of the Ta to a depth of 46~keV at half maximum. A proton energy of $E_p=318$~keV, where the thick-target yield plateau is still almost constant, was chosen for all the measurements. The target stoichiometry after the implantation reaches saturation should be equal to that of the compound Ta$_2$N$_3$ \cite{seu87, kei79}. 

\begin{figure}
  \includegraphics[angle=0,width=\columnwidth]{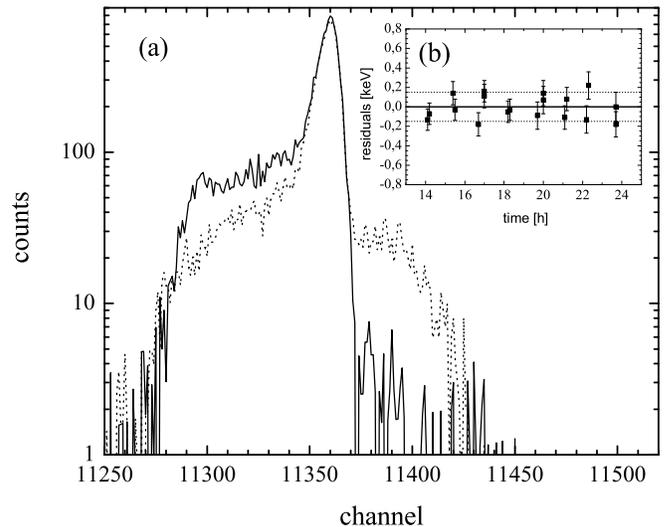}
  \caption{\label{fig16O6129}(a) Gamma-ray spectrum near the $E_\gamma=6129$~keV transition curve of $^{19}$F(p,$\alpha\gamma)^{16}$O obtained with the CaF target at $E_p = 318$~keV. The solid and dotted lines represent measurements at 0$^\circ$ and 90$^\circ$, respectively, and demonstrate the effect of the Doppler-shift at different angles. The inset (b) shows the stability of this $\gamma$-ray line versus time during the experiment in respect to the average value for both detection angles. The data are corrected for the time dependent energy shift.}
\end{figure}

In addition, a CaF target has been evaporated on a Ta backing (thickness of the CaF layer = 410 nm) and was used for calibration purpose.
This CaF target was placed on the target holder together with the $^{14}$N target and moved into the beam with the proton beam untouched in both focus and energy. This procdure was repeated before and after each run on the $^{14}$N target. Thus the gain stability of the detector/electronic system was controlled by observing the $E_\gamma=6129$~keV line of the $^{19}$F(p,$\alpha\gamma)^{16}$O reaction at $E_p=318$~keV ($E=302$~keV in the $^{19}$F + p system) at 0$^\circ$ and 90$^\circ$, respectively.
The peak position of the $E_\gamma=6129$~keV line in each calibration run was extracted from the spectrum by the determination of the centroid of the peak area. Sample spectra are shown in fig. \ref{fig16O6129}. The effect due to different contributions from unstopped $^{16}$O nuclei is clearly visible. The absence of a high energy tail suggests a strongly foreward peaked angular distribution of the $\alpha$-particles and as a consequence one observes at 0$^\circ$ besides the $\gamma$-ray emission from $^{16}$O nuclei at rest a low energy tail due to the backward motion of the $^{16}$O nuclei (solid line in fig. \ref{fig16O6129}). However, at 90$^\circ$ also a high energy tail emerges. The treatment of the background below the peak from the unshifted contributions depends on the kinematics and angular distribution of the $\alpha$-particles which is not precisely known. Different treatments of this background lead to different centroids which can amount up to 1~keV difference between a 0$^\circ$ and a 90$^\circ$ measurement. We have adopted a linear background for the 90$^\circ$ run and an error function (position and width taken from the peak) for the 0$^\circ$ case. The centroid for both runs agreed within 0.15~keV using these assumptions. 
The result of the calibration runs is that the offset in the energy calibration was constant within $\pm0.08$~keV during the course of the experiment (about 10 hours) while the relative error in the slope was $5\times10^{-5}$. Moreover, the slope shows a weak linear time dependence of about 0.025~keV/hour. All spectra were corrected for this effect. In summary, the accuracy of the energy calibration is of the order of $\pm0.15$~keV (see inset in fig. \ref{fig16O6129}).

Finally, the angles measured with the N target were chosen to be 90$^\circ$, 0$^\circ$, 105$^\circ$, 75$^\circ$, 20$^\circ$, -20$^\circ$, 116$^\circ$, -40$^\circ$, 60$^\circ$, 40$^\circ$, 82$^\circ$, 116$^\circ$, 0$^\circ$, and 90$^\circ$ in this order in time. The target holder could be placed in two positions: at +110$^\circ$ (1) and at -110$^\circ$ (2) with respect to the beam axis. The first 11 angles were obtained with the target in position 1 while the last three angles were measured in position 2. The different target positions were used to exclude possible influences of $\gamma$-ray absorption through the target holder on the determination of the energy-centroid of the peak. Moreover, the fine grid of angles allows to check experimentally a displacement of the turntable in respect to the beam spot position.

In each run or angle measurement, respectively, the centroid of five $\gamma$-ray lines was determined: the secondary transitions of the $E_x=6792$~keV state, primary ($E_\gamma=1384$~keV) and secondary of the $E_x=6172$~keV state, and primary ($E_\gamma=2385$~keV) and secondary of the $E_x=5181$~keV state. The primary transition into the $E_x=6792$~keV state was not taken into account due to too small shifts of this line. The observed $\gamma$-ray energy, $E_\gamma(\theta)$, at an angle $\theta$ in respect to the recoil direction is related to the emitted $\gamma$-ray energy , $E^0_\gamma$ (in the center-of-mass frame), by the relation:

\begin{equation}
E_\gamma(\theta)=E^0_\gamma\cdot(1+\frac{v_0}{c}P{\cdot}F(\tau)cos(\theta))
\end{equation}
where $v_0$ is the initial recoil velocity and $P$ is a correction accounting for the finite size of the Ge detector, which has been obtained by a GEANT4 simulation \cite{geant} to $P=0.987$.

\begin{figure}
  \includegraphics[angle=0,width=\columnwidth]{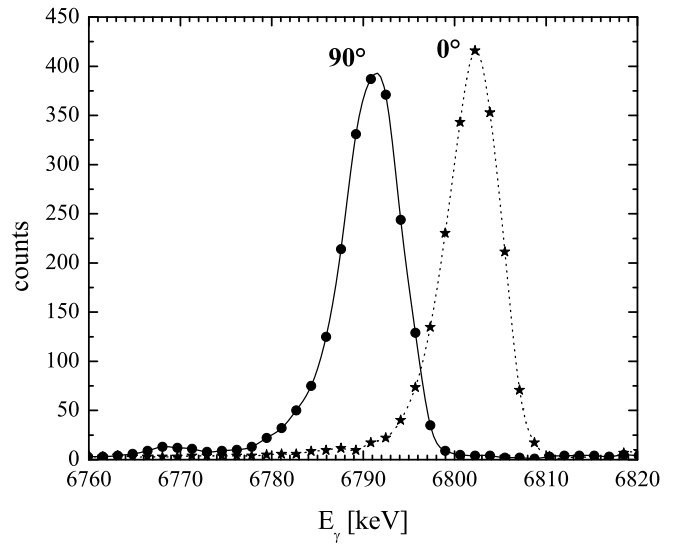}
  \caption{\label{figcompare}Gamma-ray spectra near the $E_\gamma=6792$~keV transition curve of $^{14}$N(p,$\gamma)^{15}$O obtained at $E_p = 318$~keV. The stars and the filled circles represent measurements at 0$^\circ$ and 90$^\circ$, respectively. The lines are to guide the eye.}
\end{figure}

The quality of the spectra of the present experiment are displayed in fig. \ref{figcompare} for the most important transition:  the $E_x=6792$~keV state secondary transition. The statistics of the 90$^\circ$ run (filled circles in fig. \ref{figcompare}) was improved by a factor two in the present experiment compared to previous experiments, i.e. \cite{ber01}. The background of the $E_\gamma=6792$~keV $\gamma$-ray line was - similar to the calibration runs - estimated by an error function and is nearly negligible for this line. We have not analysed the single escape lines - which in principal could be also used - because they have less statistics and higher background, which would in turn introduce larger systematic uncertainty due to the background treatment. The width of the region for the centroid determination was kept constant for all angles but shifted with the Doppler-shift of the line. 

\begin{figure}
  \includegraphics[angle=0,width=\columnwidth]{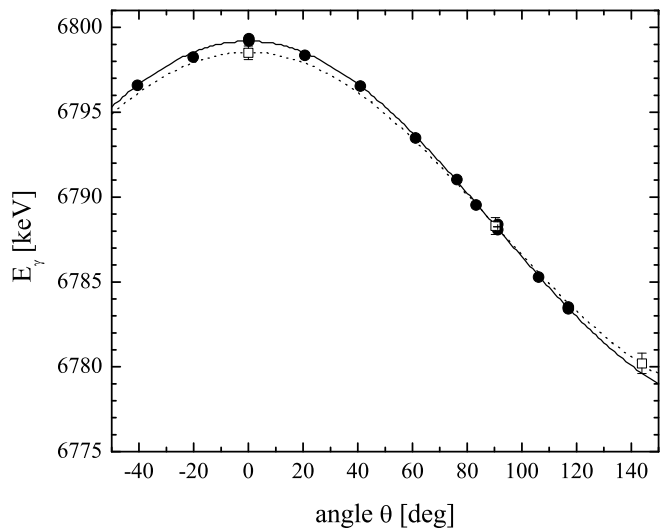}
  \caption{\label{figDScurve}Gamma-ray energy of the Doppler shifted $E_\gamma=6792$~keV line versus the detection angle $\theta$. The data points (filled circles) and the best fit ($F(\tau)=0.99$, solid line) of the present experiment are compared with the results (open squares) and the fit ($F(\tau)=0.93$, dotted line) of Bertone et al. \cite{ber01}.}
\end{figure}

The analysis of the $E_\gamma=5181$~keV line was seriously hampered by the double escape line ($E_\gamma=5150$~keV) of the 6172~keV peak. Both lines shift differently and therefore at 0$^\circ$ (116$^\circ$) the distance of both lines was only 29~keV (34~keV) complicating the background determination. These circumstances introduce additional systematic errors in the determination of the centroid for this transition. The intensity of the primary transition $\gamma$-ray lines is much higher and hence the precision with which their centroid could be analysed relative to the expected full shift is nearly the same as for the high energy lines. The positions of these lines, which are expected to show full Doppler-shift, serve as a constraint for the determination of the geometrical offset parameters of the turntable relative to the target position.

\begin{table*}
\caption{\label{Ftable}Experimental $F(\tau)$ values and corresponding lifetimes compared with literature.}
\begin{ruledtabular}
\begin{tabular}{ccccccc} 
$E_\gamma$ [keV] & $F(\tau)$ & $\chi^2$ & $\tau$ [fs] & $F(\tau)$\footnotemark[1] & $\tau$ [fs]\footnotemark[1] & $\tau$ [fs]\footnotemark[2] \\
\hline
1384\footnotemark[3] & $1.00\pm0.01$ & 0.6 & & & & \\
2375\footnotemark[3] & $0.99\pm0.01$ & 1.0 & & & & \\
5181\footnotemark[4] & $0.78\pm0.02$ & 0.4 & $8.4\pm1.0$ & $0.68\pm0.03$ & $9.67\pm1.30$ & $8.2\pm1.0$ \\
6172\footnotemark[4] & $0.99\pm0.01$ & 1.1 & $<0.77$ & $0.91\pm0.05$ & $2.1\pm1.3$ & $<2.5$ \\
6792\footnotemark[4] & $0.99\pm0.01$ & 0.7 & $<0.77$ & $0.93\pm0.03$ & $1.6\pm0.8$ & $<28$ \\
\end{tabular}
\end{ruledtabular}
\footnotetext[1]{ref. \cite{ber01}}
\footnotetext[2]{ref. \cite{ajz91}}
\footnotetext[3]{primary transition}
\footnotetext[4]{secondary transition}
\end{table*}

\begin{figure}[b]
  \includegraphics[angle=0,width=\columnwidth]{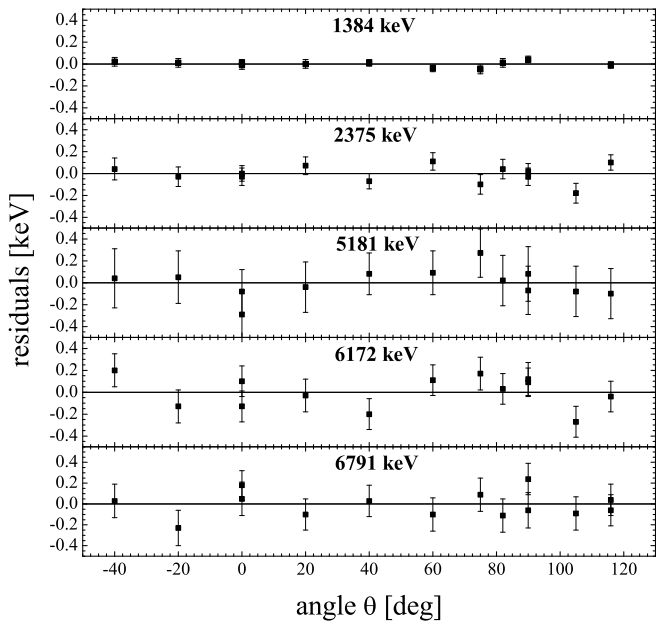}
  \caption{\label{figresiduals}Residuals of all analysed $\gamma$-ray lines for the best combined fit of the Doppler-shift attenuation according to eq. 1.}
\end{figure}

The experimental shifts $\Delta$ are obtained from the slope of a linear fit to the observed $\gamma$-ray energies versus $cos(\theta)$, with $\theta$ being the observation angle. The attenuation factor $F(\tau)=\Delta/(\Delta_{full}{\cdot}P)$ can be calculated analytically from the measured shift values $\Delta$. A correction of the observation angle $\theta$ due to the real position of the turntable was applied, where the geometrical shift parameters a and b were kept as free parameters in the fit. In a combined fit of all lines using eq. 1 the $F(\tau)$ values as well as the parameters a and b were determined, where the $F(\tau)$ values for the primary transitions were only allowed to vary within $\pm0.01$ from unity.
The best fit results in an attenuation factor of $F(\tau)=0.99\pm0.01$ for the $E_\gamma=6792$~keV $\gamma$-ray line and is displayed in fig. \ref{figDScurve} (solid line) together with the experimental data points (filled circles) for this transition.
For comparison the measured values (open squares) of \cite{ber01} and their best fit with $F(\tau)=0.93$ (dotted line) are shown: the measurements are in good agreement with the present data but the attenuation value of \cite{ber01} is excluded by more than $5\sigma$ by the dense experimental data grid of the present experiment. The conclusion from the present experiment is that the result is in good agreement with full Doppler-shift. Therefore, only a lower limit of the attenuation factor $F(\tau)$ for the $E_x=6792$~keV transition can be given and, in turn, only an upper limt for the lifetime of the associated state (see below). The results for all transitions are summarized in table \ref{Ftable} together with the calculated uncertainties and the reduced $\chi^2$. The residuals of all 5 fits are shown in fig. \ref{figresiduals}. The geometrical offset of the turntable was determined to $a=-2.6$~mm and $b=0.4$~mm within expectations.

\section{Lifetime determination}

For the lifetime extraction we followed the assumption of \cite{ber01, bis77} that the implanted N ions occupy interstitial, rather than substitutional, sites within the Ta lattice and, therefore, the density should be close to that of Ta. In this case $F(\tau)$ can be calculated by

\begin{equation}
F(\tau)=\frac{\int_{0}^{\infty}e^{-t/\tau}\overline{v(t)cos(\psi)}dt}{v_0\tau}
\end{equation}

\begin{figure}
  \includegraphics[angle=0,width=\columnwidth]{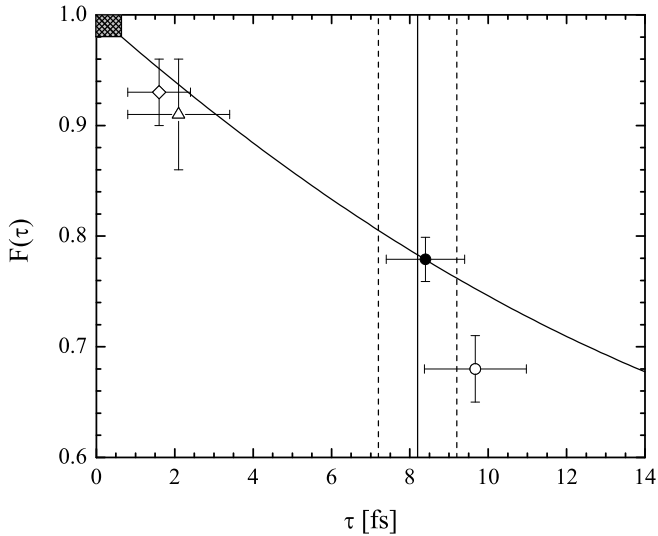}
  \caption{\label{figlifetimes}$F(\tau)$ function according to eq. 2 for $^{15}$O in Ta. The allowed region and the results of the present experiment are indicated by the shaded area in the upper left corner for $E_x=6172$ and 6792~keV and the filled circle for the $E_x=5181$~keV transition, respectively. The results of \cite{ber01} are shown as open symbols (circle = 5181~keV, triangle = 6172~keV, and diamond = 6792~keV transition). The literature value for the $E_x=5181$~keV state lifetime \cite{ajz91} is indicated by the vertical lines.}
\end{figure}

We have run the SRIM program \cite{srim} to obtain the collision details for a large number of recoiling $^{15}$O nuclei from which $\overline{v(t)cos(\psi)}$, the time dependent averaged projection of the recoil velocity distribution on the beam axis, was extracted and then fitted by a sixth order polynomial. Eq. 2 can then be solved analytically. The resulting $F(\tau)$-curve is not as steep as found in \cite{ber01} (see also open symbols compared to solid line in fig. \ref{figlifetimes}) but results in a lifetime determination of the $E_x=5181$~keV state in $^{15}$O (table \ref{Ftable} and solid point in fig. \ref{figlifetimes}) which is in excellent agreement with \cite{ajz91} (vertical lines in fig. \ref{figlifetimes}). The analysis leads to an upper limit of the lifetime of $\tau<0.77$~fs for the $E_x=6172$ and 6792~keV states, indicated by the shaded area in the upper left corner of fig. \ref{figlifetimes}. For comparison, fig. \ref{figlifetimes} also shows the values obtained by \cite{ber01} as open symbols.
If the saturation of N in the Ta was not reached completely during the implantation the resulting $F(\tau)$-curve is not as steep as in the presented case. Consequently, in the worst case of a N:Ta ratio of 1:1 the 1-$\sigma$ upper limit increases by only 2~\% for the $E_x=6172$ and 6792~keV states and to $\tau=9$~fs for the $E_x=5181$~MeV state within the given uncertainty range.

In summary, a finite lifetime for the latter states in $^{15}$O cannot be extracted from the present data. This leads to the final conclusion that there are strong doubts that the lifetime of these states can be measured by means of the Doppler-shift attenuation method. 
Moreover, ref. \cite{bis77} opens the possibility of an independent check of our analysis procedure on experimental $F(\tau)$ values. In this work the lifetime of states in $^{14}$N were obtained using the Doppler-shift attenuation method with the reaction $^{13}$C(p,$\gamma)^{14}$N at $E_p=1150$~keV. The authors also evaluated the lifetime with Monte Carlo calculations and investigated in the course of the measurements experimentally the stopping power of $^{15}$N ions in Ta. The reason to choose $^{15}$N instead of $^{14}$N was only due to the absence of any narrow $^{14}$N resonances in the range of their accelerator. Our analysis using the collision details (see above) obtained from SRIM \cite{srim} is in very good agreement with the results of \cite{bis77}, see table \ref{bistable}. We also have reproduced the experimental depth profiles of $^{15}$N ions implanted in Ta of \cite{bis77} using SRIM \cite{srim}, indicating that the stopping power calculation from \cite{srim} is sufficiently precise for the present experiment.

\begin{table}
\caption{\label{bistable}Lifetime determination with the present analysis method from experimental $F(\tau)$ values \cite{bis77} obtained by the reaction $^{13}$C(p,$\gamma)^{14}$N compared with the results from \cite{bis77}. The very good agreement of the results demonstrate the validity of the present analysis for the lifetime $\tau$.}
\begin{ruledtabular}
\begin{tabular}{cccc} 
level in $^{14}$N & \multicolumn{2}{c}{ref. \cite{bis77} Bister et al.} & present \\
$E_x$ [keV] & $F(\tau)$ & $\tau$ [fs] & $\tau$ [fs] \\
\hline
3948 & $0.894\pm0.005$ & $8.4\pm0.4$ & $8.1\pm0.4$ \\
5690 & $0.82\pm0.08$ & $16\pm8$ & $14\pm8$ \\
\end{tabular}
\end{ruledtabular}
\end{table}

\section{Summary and conclusion}

The present experiment has improved the statistical as well as systematic uncertainties in the determination of the Doppler-shift attenuation of three $\gamma$-ray transitions in $^{15}$O: 5181, 6172 and 6792~keV. Our upper limits on the lifetimes of the $E_x=6172$ and 6792~keV is only marginaly in agreement with the conclusion of \cite{ber01}, while the experimental data of \cite{ber01} are not in contradiction with the present data.

The upper limit for the lifetime of the $E_x=6792$~keV subthreshold state in $^{14}$N(p,$\gamma)^{15}$O converts to a lower limit of the $\gamma$-width of this state of $\Gamma_\gamma>0.85$ eV.
This result is in good agreement with the value found in \cite{for04} with $\Gamma_\gamma=0.8\pm0.4$ eV from R-matrix analysis of the S~factor curve of the ground state transition in $^{14}$N(p,$\gamma)^{15}$O. 
The S~factor extrapolation from the R-matrix fit of \cite{run05} indicates a value of $\Gamma_\gamma\approx1.6$ eV not in contradiction with the present result, but in disagreement with \cite{for04} (Note: the R-matrix fit of \cite{run05} was obtained by omitting the high energy data of Schr{\"o}der et al. \cite{sch87}). 
In a Coulomb excitation experiment \cite{yam04} a width of $\Gamma_\gamma=0.95^{+0.6}_{-0.95}$ eV was reported in agreement with the present result. Hence, the statements in \cite{for04, imb05, run05} concerning the ground state transition in $^{14}$N(p,$\gamma)^{15}$O are essentially unchanged, but further conclusions need to await a more detailed R-matrix study. Moreover, further investigations for the ground state S~factor of the reaction $^{14}$N(p,$\gamma)^{15}$O are still necessary. Such an experiment \cite{tra08} was performed recently, i.e. a precision measurement of the ground state transition above the $E_R=259$~keV resonance with a Clover detector, reducing the necessary summing correction by one order of magnitude compared to previous direct studies \cite{for04, run05}. The results \cite{mar08} indicate again a very good agreement with the present result and the $\Gamma_\gamma$ of Formicola et al. \cite{for04}.

\begin{acknowledgments}
The authors thank H.W.~Becker and J.N.~Klug (RUBION Bochum, Germany) for their support with the 500~kV accelerator and Z.~F\"ul\"op (ATOMKI Debrecen, Hungary) for fruitful comments on the manuscript.
\end{acknowledgments}

\bibliography{doppler} 

\end{document}